\documentclass[fleqn,twoside]{article}
\usepackage{espcrc2}
\usepackage{epsfig}

\usepackage{graphicx}
\usepackage[figuresright]{rotating}

\newcommand{\munuebar}{\rm{\mu_{\nuebar}}}
\newcommand{\mub}{\rm{\mu_B}}

\newcommand{\nuebar}{\bar{\nu_e}}
\newcommand{\nue}{\nu_e}
\newcommand{\munu}{\mu_{\nu}}
\newcommand{\gammanu}{\Gamma_{\nu}}

\newcommand{\AmS}{{\protect\the\textfont2
  A\kern-.1667em\lower.5ex\hbox{M}\kern-.125emS}}

\title{
{\normalsize \hfill AS-TEXONO/03-08 \\
\hspace*{1cm} \hfill \today\\}
Low Energy Neutrino Physics at the Kuo-Sheng
Reactor Neutrino Laboratory in
Taiwan}

\author{Henry T. Wong \address{Institute of Physics, Academia Sinica, 
			Taipei 11529, Taiwan}
	}

\begin{document}

\begin{abstract}
A laboratory has been constructed by 
the TEXONO Collaboration at the Kuo-Sheng Reactor Power
Plant in Taiwan to study low energy neutrino physics.
The facilities of the laboratory are described.
A limit on the neutrino magnetic moment of
$\rm{ \munuebar < 1.3 \times 10^{-10} ~ \mub}$
at 90\%  confidence level has been achieved from
measurements with a high-purity germanium detector.
Other physics topics, as well as 
the various R\&D program, are surveyed.

\vspace{1pc}
\end{abstract}

\maketitle

\section{Introduction}
\label{intro}
The
TEXONO\footnote{{\bf T}aiwan {\bf EX}periment {\bf O}n 
{\bf N}eutrin{\bf O}} 
Collaboration\footnote{Home Page - 
http://hepmail.phys.sinica.edu.tw/$\sim$texono}
has been built up since 1997 to
pursue an experimental
program in Neutrino and Astroparticle Physics~\cite{program}.
The ``flagship'' program is on reactor-based 
low energy neutrino physics
at the Kuo-Sheng (KS) Power Plant in Taiwan.
The KS experiment is the first 
large-scale particle 
physics experiment in Taiwan.
The TEXONO Collaboration
is the first research collaboration
among scientists from Taiwan and China~\cite{sciencemag}.

Results from recent neutrino experiments
strongly favor neutrino oscillations
which imply neutrino masses and
mixings~\cite{nu02}.
Their physical origin and experimental consequences
are not fully understood.
There are strong motivations for
further experimental efforts to
shed light on these fundamental questions
by probing standard and anomalous neutrino properties
and interactions~\cite{valle}. 
The results
can constrain theoretical models
necessary to interpret
the future precision data $-$ or may yield surprises
which have been the characteristics of the field.
In addition, these studies
could also explore new detection channels to
provide new tools for future investigations.

\section{Kuo-Sheng Neutrino Laboratory}
\label{kslab}

The ``Kuo-Sheng Neutrino Laboratory''
is located at a distance of 28~m from the core \#1
of the Kuo-Sheng Nuclear Power Station
at the northern shore of Taiwan~\cite{program}.
A multi-purpose ``inner target'' detector space of
100~cm$\times$80~cm$\times$75~cm is
enclosed by 4$\pi$ passive shielding materials
with a total weight of 50 tons.
Different detectors can be placed in the
inner space for the different scientific goals.
The detectors are read out by a versatile
electronics and data acquisition systems~\cite{eledaq}
based on 16-channel, 20~MHz, 8-bit
Flash Analog-to-Digital-Convertor~(FADC) modules.
The readout allows full recording of all the relevant pulse
shape and timing information for as long as several ms
after the initial trigger.
The reactor laboratory is connected via telephone line to
the home-base laboratory, where remote access
and monitoring are performed regularly. Data are stored
and accessed with a cluster of multi-disks arrays
each with 800~Gbyte of memory.

\begin{figure}[hbt]
\center
\epsfig{file=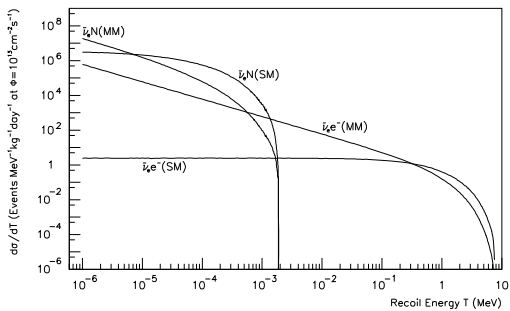,width=7cm}
\caption{
Differential cross section showing the
recoil energy spectrum in
$\nuebar$-e and coherent $\nuebar$-N
scatterings,
at a reactor neutrino flux of
$\rm{10^{13}~cm^{-2} s^{-1}}$,
for the Standard Model (SM) processes and
due to a neutrino
magnetic moment (MM) of 10$^{-10}~\mub$.
}
\label{spectra}
\end{figure}

The measure-able nuclear and electron recoil spectra
due to reactor $\nuebar$ are depicted in Figure~\ref{spectra},
showing the effects due to
Standard Model [$\rm{\nuebar e^- (SM) }$] and
magnetic moment [$\rm{\nuebar e^- (MM) }$] 
in $\nuebar$-electron
scatterings~\cite{sigmanue}, as well as 
in neutrino coherent scatterings on the nuclei
($\rm{\nuebar N (SM) }$ and $\rm{\nuebar N (MM) }$, respectively).
The uncertainties in the
low energy part of the reactor neutrino spectra
require that
experiments to measure $\rm{\sigma [ \nuebar e^- (SM) ] }$
should focus on higher electron recoil
energies (T$>$1.5~MeV), while
MM searches should base
on measurements with T$<$100~keV~\cite{sensit}.
Observation of $\rm{\rm{\nuebar N (SM) } }$ would
require detectors with sub-keV sensitivities.

Accordingly, data taking were optimized 
with these strategies.
An ultra low-background high purity germanium (ULB-HPGe)
detector was used for Period I (June 2001 till May 2002)
data taking, while 186~kg of CsI(Tl) crystal scintillators
were added in for Period II (Jan 2003 till Sept 2003).
Both detector systems operate in parallel
with the same data acquisition system
but independent triggers.

\section{Neutrino Magnetic Moment Searches with Germanium Detector}
 
The ULB-HPGe 
is surrounded by NaI(Tl) and CsI(Tl) crystal
scintillators as anti-Compton detectors, 
and the whole set-up
is further enclosed by another 3.5~cm of OFHC copper 
blocks, and housed in a radon shield.
After suppression of cosmic-induced
background, anti-Compton vetos and convoluted events
by pulse shape discrimination,
a background level at 20~keV
at the range of 1~keV$^{-1}$kg$^{-1}$day$^{-1}$
and a detector threshold of 5~keV are  achieved.
These are the levels comparable to underground Dark Matter
experiment.
Comparison of the measured spectra for 4712/1250 hours of
Reactor ON/OFF data in Period I~\cite{munupaper}  
shows no excess
and limits of the neutrino magnetic moment
$\rm{ \munuebar < 1.3(1.0) \times 10^{-10} ~ \mub}$
at 90(68)\% confidence level (CL) were derived.

\begin{figure}[hbt]
\center
\epsfig{file=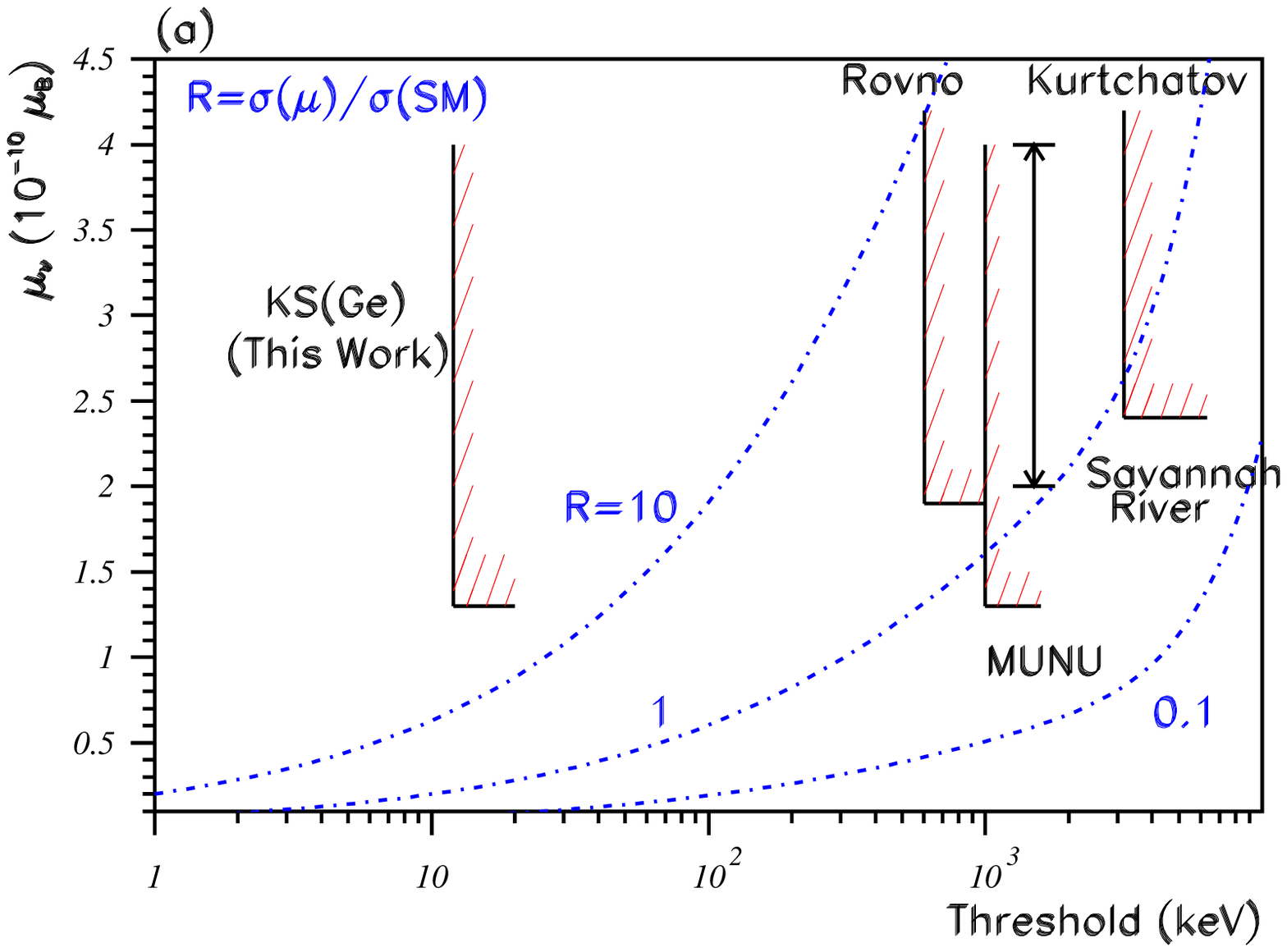,width=6cm}
\epsfig{file=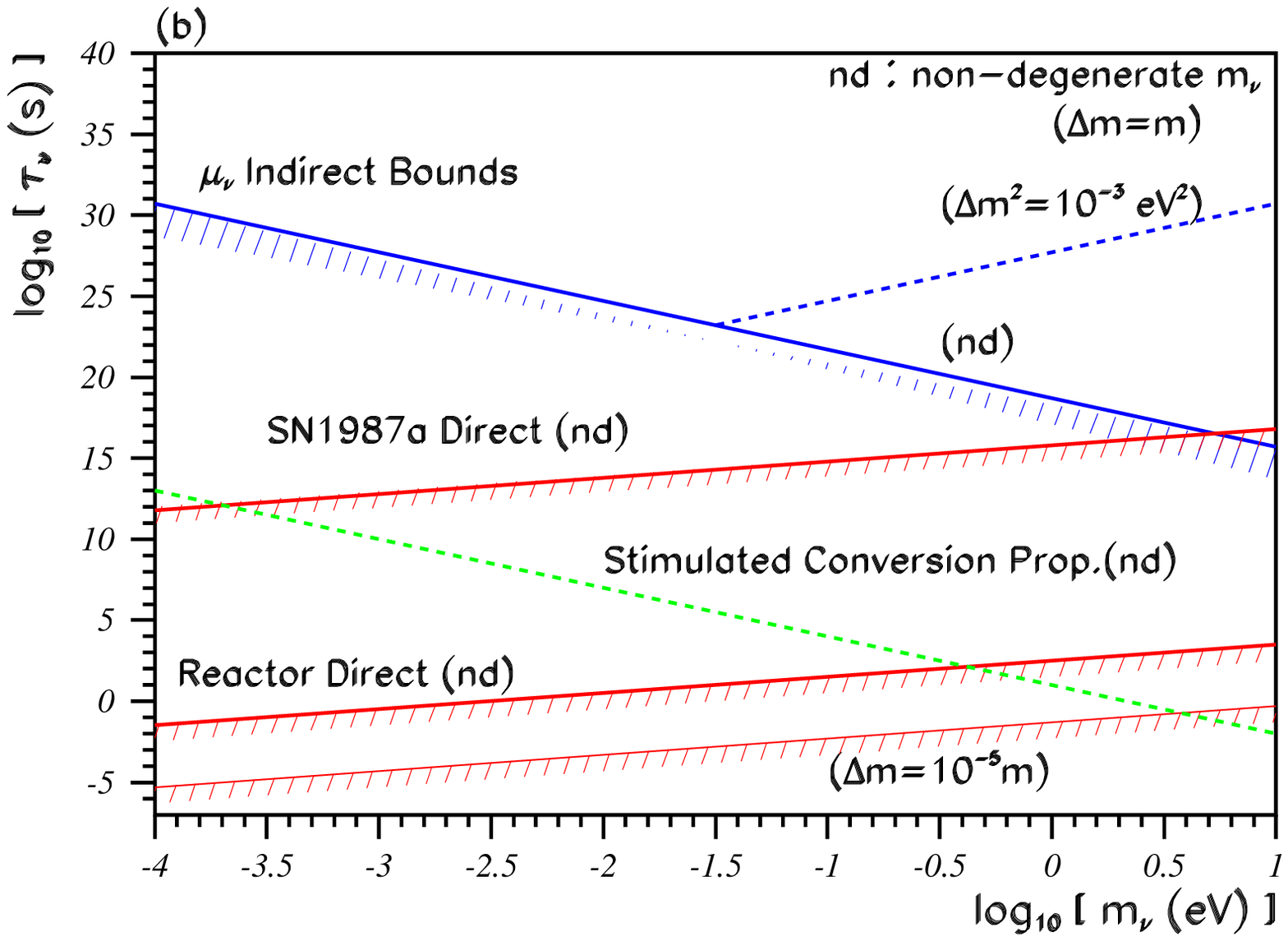,width=6cm}
\caption{
Summary of the KS Period-I results in
(a) the searches of neutrino
magnetic moments with reactor neutrinos,
and
(b) the bounds of
neutrino radiative decay lifetime.
}
\label{summaryplots}
\end{figure}

Depicted in Figure~\ref{summaryplots}a is the
summary of the results in $\rm{\munuebar}$ searches
versus the achieved threshold
in reactor experiments.
The dotted lines denote the
$\rm{R = \sigma ({\mu}) / \sigma (SM) }$ ratio at a
particular (T,$\rm{\munuebar}$).
The KS(Ge) experiment has
a much lower physics threshold of 12~keV compared
to the other measurements.
The large R-values
imply that the KS results
are robust against the uncertainties in the
SM cross-sections.
The neutrino-photon couplings
probed by $\munu$-searches in
$\nu$-e scatterings are related to
the neutrino radiative decays ($\gammanu$)~\cite{rdk}.
Indirect bounds on $\gammanu$ can be inferred
and are displayed in Figure~\ref{summaryplots}b
for the simplified scenario where a single channel
dominates the transition. It corresponds to
$\rm{
\tau_{\nu} m_{\nu} ^ 3 > 2.8(4.8) \times 10^{18} ~  eV ^3 s
}$ at 90(68)\% CL in the non-degenerate case.
It can be seen that $\nu$-e scatterings give much more
stringent bounds than the direct approaches.

\section{Other Research Subjects}

The KS data with ULB-HPGe are the lowest threshold data
so far for reactor neutrino experiments, and therefore 
allow the studies of several new and more speculative topics.
Nuclear fission at reactor cores also produce 
electron neutrino ($\nue$) through the 
production of unstable isotopes, such as
$^{51}$Cr and $^{55}$Fe, via neutron capture.
The subsequent decays of these isotopes by electron capture
would produce mono-energetic $\nue$. 
A realistic neutron transfer simulation
has been carried out to estimate the flux.
Physics analysis on the $\munu$ and $\gammanu$ of $\nue$
will be performed, while the potentials for other 
physics applications will be studied.
In additional, potentials
for an {\it inclusive} analysis of
the anomalous neutrino interactions with matter,
as well as studies on neutrino-induced
nuclear transitions will be pursued. 

Period II data taking includes an addition of
an array of 186~kg of CsI(Tl) crystals.~\cite{proto},
each module being 2~kg in mass and 40~cm in length.
The physics goal is to measure
the Standard Model neutrino-electron
scattering cross sections, and thereby
to provide a measurement of $\rm{sin ^2 \theta _W}$
at the untested MeV range.
The strategy~\cite{sensit} is
to focus on data at high ($>$2~MeV) recoil
energy where the uncertainties due to
the reactor neutrino spectra are small.
The large target mass compensates the
drop in the cross-sections at high energy.

\begin{figure}[hbt]
\center
\epsfig{file=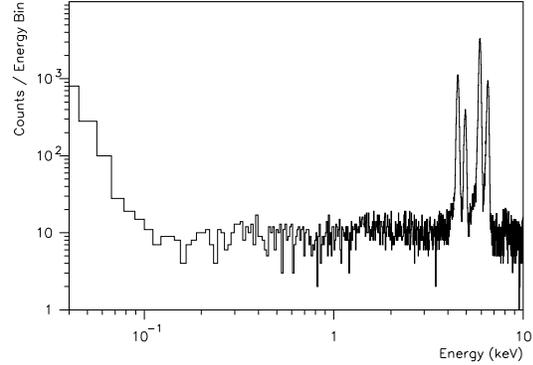,width=7cm}
\caption{
Energy spectrum 
with the ULE-HPGe,
indicating a detector threshold lower
than 100~eV,
The peaks are from an $^{55}$Fe 
source and from back-scattering with Ti.
}
\label{lege}
\end{figure}

In addition, various R\&D projects~\cite{program} 
are pursued in parallel to the
KS reactor neutrino experiment.
In particular, a prototype 
ultra-low-energy germanium (ULE-HPGe)
detector of 5~g mass is being studied,
with potential applications on
Dark Matter searches  and
neutrino-nuclei coherent scatterings.
A hardware energy threshold of better than
100~eV has been achieved, 
as illustrated in Figure~\ref{lege}.
The ULE-HPGe
is placed inside the shieldings
at the KS laboratory where the goal 
will be to perform 
the first-ever background studies
at the sub-keV energy range.
It is technically feasible to build an array of
such detectors to increase the target size to
the 1~kg mass range.

The TEXONO research program is supported by
the National Science Council, Taiwan and
the National Science Foundation, China,
as well as by the operational funds from
the collaborating institutes.

\end{document}